\documentclass{aa}
\usepackage{graphicx,amsmath}
\usepackage{txfonts,color}

\usepackage{natbib} 
\bibpunct{(}{)}{;}{a}{}{,}

\begin{document}
\title{Stray-light contamination and spatial deconvolution of slit-spectrograph observations}

   \author{C. Beck\inst{1,2} \and R. Rezaei\inst{3} \and D. Fabbian\inst{1,2}}
   \titlerunning{Stray-light contamination and spatial deconvolution of spectrograph observations}
   \authorrunning{C. Beck, R. Rezaei \& D. Fabbian}  

   \institute{Instituto de Astrof\'{\i}sica de Canarias
         \and Departamento de Astrof{\'i}sica\\
     \and Kiepenheuer-Institut f\"ur Sonnenphysik\\
        }
 
\date{Received xxx; accepted xxx}
\abstract{Stray light caused by scattering on optical surfaces and in the
  Earth's atmosphere degrades the spatial resolution of observations. Whereas
  post-facto reconstruction techniques are common for 2D imaging
  and spectroscopy, similar options for slit-spectrograph data are rarely
  applied.}{We study the contribution of stray light to the two channels of
  the POlarimetric LIttrow Spectrograph (POLIS)  at 396\,nm and 630\,nm as an
  example of a slit-spectrograph instrument. We test the performance of
  different methods of stray-light correction and spatial deconvolution to
  improve the spatial resolution post-facto.}{We model the stray light as
  having two components: a spectrally dispersed component and a ``parasitic''
  component of spectrally undispersed light caused by scattering inside the
  spectrograph. We used several measurements to estimate the two
  contributions: a) observations with a (partly) blocked field of view (FOV),
  b) a convolution of the FTS spectral atlas, c) imaging of the spider
  mounting in the pupil plane, d) umbral profiles, and e)  spurious
  polarization signal in telluric spectral lines. The measurements with a
  partly blocked FOV in the focal plane allowed us to estimate the
  spatial point spread function (PSF) of POLIS and the main spectrograph of
  the German Vacuum Tower Telescope (VTT). We then used the obtained PSF
  for a deconvolution of both spectroscopic and spectropolarimetric data and
  investigated the effect on the spectra.}{The parasitic contribution can be directly and accurately determined for POLIS, amounting to about 5\,\% (0.3\,\%) of the (continuum) intensity at 396\,nm (630\,nm). The spectrally dispersed stray light is less accessible because of its many contributing sources. We estimate a lower limit of about 10\,\% across the full FOV for the dispersed stray light from umbral profiles. In quiet Sun regions, the stray-light level from the  close surroundings ($d< 2^{\prime\prime}$) of a given spatial point is about 20\,\%. The stray light reduces to below 2\% at a distance of 20$^{\prime\prime}$ from a lit area for both POLIS and the main spectrograph. The spatial deconvolution using the PSF obtained improves the spatial resolution and increases the contrast, with a minor amplification of noise.}{A two-component model of the stray-light contributions seems to be sufficient for a basic correction of observed spectra. The instrumental PSF obtained can be used to model the off-limb stray light, to determine the stray-light contamination accurately for observation targets with large spatial intensity gradients such as sunspots, and also to improve the spatial resolution of observations post-facto.}
\maketitle
\section{Introduction}
The importance of stray-light contributions to observed spectra and images was
realized very early on and continues to be a problem today \citep[see, e.g.,][]{henoux1969,mattig1971,martinezpillet1992,chae+etal1998,wedemeyer2008,deforest+etal2009,mathew+etal2009}. The intrinsically limited optical
quality of the reflective surfaces of mirrors or of the glass of lenses leads
to scattering of photons in the light path. In addition, scattering during the
passage through the Earth's atmosphere by, for instance, dust particles
spreads the light from the full solar disk to any point of an observed
restricted field of view (FOV). Depending on the wavelength and the target of
the observations, the stray light can amount to a significant fraction of the
observed intensity and can have a strong impact on the final results of the data analysis. Therefore, many inversion codes that determine physical quantities from observed spectra, such as the SIR code \citep{cobo+toroiniesta1992}, include an explicit treatment of this contamination by using a separately provided stray-light profile \citep[see also, e.g.,][]{orozco+etal2007,orozco+etal2007a}.

For analysis techniques that make direct use of the observed spectra or of
broad-band images, the stray light has to be dealt with in advance. This
problem had to be tackled in the context of the accurate determination of
sunspot intensities to derive their temperature at continuum forming layers
\citep{kneer+mattig1968,maltby+mykland1969,maltby1970,mattig1971,tritschler+schmidt2002}
or in the determination of the continuum contrast of solar granulation
\citep{mathew+etal2009,wedemeyer+etal2009}. It is also important in studies of
solar chemical abundances that seek to derive consistent results from different spectral lines \citep{asplund+etal2009,fabbian+etal2010}.

The stray-light contribution is often decomposed in different sources: 1.~the
atmospheric stray light caused by large-scale scattering in the Earth's
atmosphere, with a slow temporal evolution because of the change of
air-mass in the light path during the day; 2.~the ``blurring'', i.e., the
rapidly changing  amount of stray light caused by the fluctuations of the
refractive index in the Earth's atmosphere (which are nowadays commonly
referred to as ``seeing''); and 3.~instrumental stray light, i.e., scattering
in the telescope/instrument optics or (static) stray-light effects of the
telescope/instrument caused by the geometry of the finite aperture and the
possible presence of (central) obscurations in the light path
\citep{zwaan1965,staveland1970,mattig1983,martinezpillet1992,wedemeyer2008}.
For space-based observations, naturally, the first two points are absent. 

The stray-light contamination is quantified in the ``spread function'', whose
large-scale and small-scale contributions are usually approximated as
analytical functions of varying shapes, e.g., Gaussian or Lorentzian, or with
combinations of a number of them
\citep{mattig1971,wedemeyer2008,deforest+etal2009,mathew+etal2009}. For
describing the purely instrumental effects without temporal dependence and in
night-time astronomy, the label ``spatial point spread function'' (PSF) is commonly used because the function describes the spatial shape that the light from a point source (as a star can be approximated to be) would
attain after passing through the optical system. 

To determine the exact shape of the spread function, either theoretical
calculations of the optical systems or observations with an only partly
illuminated FOV are used. The theoretical calculations are usually limited to
the time-invariant effects of the optics, while for the varying contributions from, e.g., the seeing only some statistically averaged effects may be
assumed. For determining the spread function in
the solar case, suitable observations can be obtained in different ways
\citep{mattig1971,briand+etal2006,wedemeyer2008,deforest+etal2009}, namely by
planetary transits in front of the Sun, by observations near and off the solar
limb, or by artificial occulters in the light path. Planetary transits are
actually the best suited because they provide the perfect reference of an
intrinsically zero light level, but the diameters of the planets are usually
smaller than the extent of the PSF and thus do not allow to determine the far
wings of the PSF; moreover, transits are rare events. Thus, the
radial variation of the residual intensity beyond the solar limb
(``aureole'') has often been used to determine the spread function of
ground-based telescopes. Some of the older
observations, however, have to be treated with some care because they were done with telescopes without an aperture field stop; i.e., the
telescopes created a full-disk solar image in the focal plane without a
restriction to a limited FOV. This implies that there was a large contribution from the full-disk stray light in those observations. Additionally, these data were usually obtained without active compensation for seeing effects by a
correlation tracker \citep{ballesteros+etal1996} or adaptive optics
\citep[AO,][]{vdluehe+etal2003,scharmer+etal2003,rimmele2004a}, therefore
mixing seeing and atmospheric scattering. \citet{mattig1983} found that the stray light from near the solar limb (up to one solar radius above it) is dominated by
instrumental and seeing effects. The atmospheric scattering only becomes
dominant at larger distances to the Sun or in the case of coronagraphs, where
the illumination from the solar disk is blocked inside the telescope
optics. \citet{briand+etal2006} show that the determination of the spread
function from the aureole does not necessarily provide a good measure of the instrumental stray light because the telescope/instrument optics are only partly illuminated near the limb, which yields a different amount of scattering than for the full illumination on disk center. 

Even if the static instrumental part of the PSF is known from either
calculations or measurements, the time-variant part caused by the Earth's
atmosphere is usually unknown in the case of ground-based observations. For
intrinsically 2D observations, whether broad-band imaging or
narrow-band spectroscopy or spectropolarimetry with suitable 2D instruments \citep[see, e.g.,][]{tritschler+etal2002,mikurda+etal2006,cavallini2006,puschmann+etal2006,scharmeretal08,beck+etal2010}, the instantaneous PSF can be estimated using series of rapid short-exposed images. Two different approaches, speckle imaging \citep[e.g.,][]{vdluehe1993} and blind deconvolution \citep[e.g.,][and references therein]{vannortetal05}, then allow a post-facto reconstruction of the most probable original solar image that lies behind the observed image series. The reconstruction algorithms are, however, only capable of recovering the instantaneous PSF within some limits and are also partially blind to the static parts of the PSF \citep[][]{scharmer+etal2010}. 

For slit-spectrograph observations, at any given moment of time, only a
1D slice of the solar surface is available, whereas the PSF is a 2D function that therefore also introduces stray light from locations
that are not covered by the slit in that moment. Thus, even if both the static
and the time-variant parts of the PSF are known, it is not straightforward to
use this information to improve the spatial resolution of slit-spectrograph
data post-facto. \citet{keller+johannesson1995} and
\citet{suetterlin+wiehr2000} introduced a method similar to a speckle
deconvolution, requiring a fast series of slit-spectra. In our
contribution, we tested the application of a post-facto correction of ``standard'' slit-spectrograph data for the static part of the instrumental PSF as obtained from a dedicated set of measurements. 

The stray-light contribution becomes significant when the light level in
observations is low, as for observations of sunspots, near the solar
limb, or for spectroscopy of very deep spectral lines such as H$\alpha$
or \ion{Ca}{ii} H. For a recent study of the center-to-limb variation (CLV) of
\ion{Ca}{ii} H spectra taken with the POlarimetric LIttrow Spectrograph
\citep[POLIS;][]{beck+etal2005b} at the German Vacuum Tower Telescope
\citep[VTT,][]{schroeter+soltau+wiehr1985}, we thus had to investigate the
stray-light contamination in detail. During the CLV run and some of the subsequent observational campaigns, we obtained various data sets for characterizing the stray-light contamination for both POLIS and the main spectrograph of the VTT. Because the measurements also allow one to estimate the spatial PSF of the instrument, we additionally investigated the option of using the known PSF for a spatial deconvolution of the slit-spectrograph observations. 

In Sect.~\ref{stray_model_gen}, we outline the basic approaches to
  correcting stray light based on the observed spectra and generic
  estimates at first (Sect.~\ref{stray_model}), and then explicitly taking the instrumental PSF into account (Sect.~\ref{deconv_sect}). The
  measurements used to derive the generic stray-light estimates and the PSF
  are described in Sect.~\ref{meas_sect}. Section \ref{deconv} shows examples
  of the application of both the stray-light correction and the deconvolution
  to a few data sets. The application to observational data 
  allowed us to cross-check and improve the stray-light estimates. Section
  \ref{alpha_true} shows the relation between the generic stray-light
  correction and the explicit use of the PSF for a few data sets. Our
  findings are discussed in Sect.~\ref{sect_disc}. Section \ref{concl}
  presents the conclusions.
\begin{figure}
\centerline{\resizebox{8cm}{!}{\includegraphics{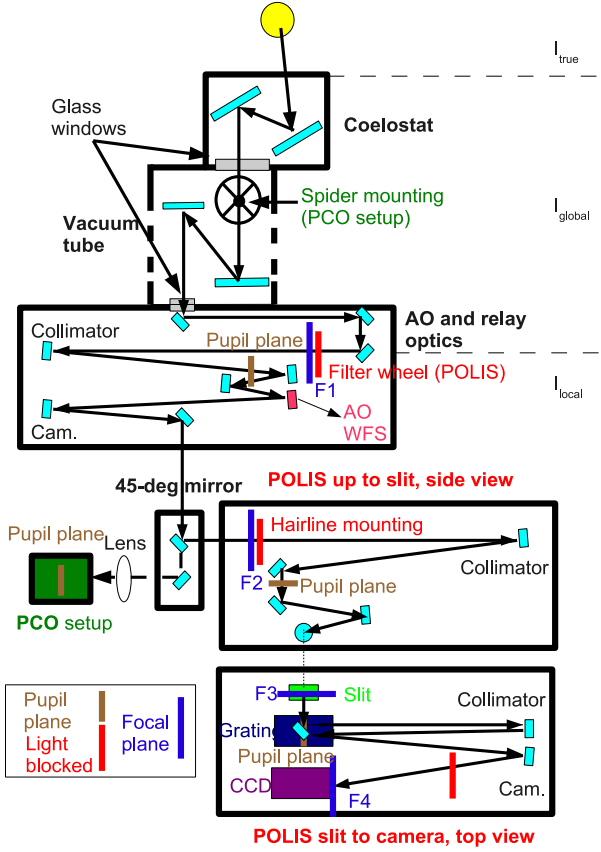}}}
\caption{Sketch of the light path at the VTT (drawing not to scale). Focal planes are denoted by {\em blue rectangles}, pupil planes by {\em brown} ones. {\em Red rectangles} denote an artificial blocking of the light path. The 45-deg mirror behind the relay optics was used to direct the light to either the PCO setup or towards POLIS. The design of POLIS is shown in side view up to the location of the slit ({\em upper panel} at lower right) and in top view for the rest of the optics ({\em lower panel} at lower right). \label{fig1}} 
\end{figure}
\section{Theoretical description of the stray-light problem\label{stray_model_gen}}
The transfer through the optical system composed of the (fluctuating) atmosphere of the Earth, the telescope, and finally the instrument modifies the real intensity distribution $I_{\rm true}$ and yields instead an observed intensity $I_{\rm obs}$ that is related to $I_{\rm true}$ by
\begin{equation}
I_{\rm obs} (t) = {\rm PSF}(t) \otimes I_{\rm true}(t) \,, \label{eq_img}
\end{equation}
where PSF(t) is the instantaneous optical point spread function and $\otimes$
denotes a convolution. 

The three contributions to the PSF are explained in the introduction
above. For observations with real-time correction by an AO system, the
atmospheric part is already (partially) corrected for, leaving the large-scale
atmospheric scattering and the static contributions from optics. The PSF
spills the light from one location into its surroundings and vice versa, which
adds additional intensity to all points in the FOV via false light originating
in other locations in the FOV, the so-called stray-light
contamination. To apply a correction for the stray-light contamination, we first derive a direct approach based on only the observed spectra and a generic estimate of the stray-light fraction in a simplified approach in the next section, and a more explicit model taking the instrumental PSF into account in Sect.~\ref{deconv_sect}.
\subsection{Modeling of stray-light contributions \label{stray_model}}
For the direct correction of the stray-light contamination, we model the intensity spectrum $I_{\rm obs} (\lambda,x,y)$ on a certain CCD row in the focal plane of the POLIS CCD cameras (focal plane F4, see Fig.~\ref{fig1}) as
\begin{eqnarray}
I_{\rm obs} (\lambda,x,y) = I_{\rm true} (\lambda,x,y) + \alpha_1 \, I_{\rm global}(\lambda) + \alpha_{2} \, I_{\rm  local}(\lambda)+\mbox{const.},  \label{eq_stray}\end{eqnarray}
with
\begin{eqnarray}
\mbox{const. } = \mbox{dc} + \mbox{parasitic light} = \mbox{dc} + \alpha_3 \, <I_{\rm local}>\; \neq f(\lambda)\,,  \label{eq_para}
\end{eqnarray}
where ``dc'' denotes the dark current of the CCD. $I_{\rm true} (\lambda,x,y)$
is the spectrum that emerges from the spatial location $(x,y)$ on the Sun that
corresponds to the CCD pixel row $y$ when the slit is located at position $x$
on the solar image. The average intensity spectrum before the first focal plane F1 is denoted by $I_{\rm global}(\lambda)$; the full solar disk contributes to it. Because of the CLV of the solar intensity, $I_{\rm global}(\lambda)$ will be dominated by
contributions from the disk center and its surroundings, and its shape is
independent of the telescope pointing. This term corresponds to the
integration over the full solar disk weighted by a spread function, as done
in, e.g., \citet{mattig1971}. It can be assumed that the stray-light level
proportional to $I_{\rm global}(\lambda)$ is uniform across the FOV, with no
direct relation to the solar location $(x,y)$ where $I_{\rm true}
(\lambda,x,y)$ originates. Scattering by any of the optical elements in front
of F1 will only need to produce slight beam deviations from the optical axis
to spill light from the full aperture to each location on the focal plane because of the long path length before F1 is reached. The factor $\alpha_1$ then refers to all nine optical elements in front of F1 (7 mirrors and 2 glass windows, see the upper part of Fig.~\ref{fig1}).

The average intensity spectrum behind F1, $I_{\rm local}(\lambda)$, comes only from a restricted FOV and not from the full solar disk because of the field stop in the filter wheel near F1. The shape of $I_{\rm
  local}(\lambda)$ depends on the telescope pointing. For the CLV
observations, we used an aperture field stop with a diameter of 40 mm
($\approx$180$^{\prime\prime}$). The diameter of the field stop in F2 inside of
POLIS is similar to the one of the field stop in F1. The stray light
proportional to $I_{\rm local}(\lambda)$ actually is made up of two contributions:
\begin{eqnarray}
\alpha_2\,I_{\rm local}(\lambda) &=& \alpha_4\, \frac{\int_{A} I(\lambda) \,dA }{\int_{A} dA}  \nonumber\\
 &+& < {\rm PSF_{\rm instr}}(\Delta x, \Delta y) \times I (\lambda,\Delta x,
 \Delta y) >_{(\Delta x,\Delta y)} \;, \label{eq_psf}
\end{eqnarray} 
where $A$ is the area of the field stop in F1, and $(\Delta x,
\Delta y)$ the distance between $(x,y)$ and all other points inside the FOV
that pass through the field stop. 

The first term is  caused by the fact that the optical elements between F1
and F4 will uniformly mix the light across the full aperture of the field stop
(as was the case for the optics upfront of F1), especially because there
are three additional pupil planes in-between. The second term depends on the spatial PSF of POLIS that introduces a strongly field-dependent stray-light contribution to the spectrum from locations close to $(x,y)$ when $(\Delta x,\Delta y)$ is smaller than about $10^{\prime\prime}$ (see Sect.~\ref{sect_psf}). ${\rm PSF_{\rm instr}}$ only refers to the optics behind F1. We assume that both contributions on the right-hand side of Eq.~(\ref{eq_psf}) can be modeled as being proportional to the same profile $I_{\rm local}(\lambda)$ because in quiet Sun regions averaging over a small (a few arcsecs$^2$) or large (several ten arcsecs$^2$) area inside the local FOV will yield a similar profile. The factor $\alpha_2$  then refers to all optics between F1 and F4 (for POLIS: 15 mirrors, modulator, grating, interference filter, and the polarizing beam splitter for the 630\,nm channel). The stray-light contributions proportional to $I_{\rm global}(\lambda)$ and $I_{\rm local}(\lambda)$ are both assumed to pass through a grating, and thus to be spectrally dispersed. 

\begin{figure}
\centerline{\resizebox{8.cm}{!}{\includegraphics{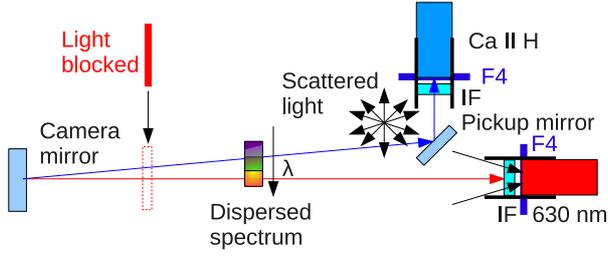}}}
\caption{Design of POLIS close to the CCD cameras in a side view (not to scale). ``IF'': order selecting interference filters. The {\em thin red rectangle} indicates the location where the light path was blocked for measuring the parasitic stray light.\label{fig10}}
\end{figure}

The ``stray light'' without wavelength dependence, the constant term in
Eq.~(\ref{eq_stray}), again contains two contributions: the usual dark current (dc) of the CCD cameras and a second contribution that
corresponds to light that reaches the CCD without passing through the grating, called ``parasitic'' light. The source of the parasitic light photons
in the case of POLIS is indicated schematically in Fig.~\ref{fig10}, which shows the optical design
of POLIS just in front of the CCD cameras. In the light path behind the
grating, a small pick-up mirror is used to deflect the blue part (indicated by
a blue ray in the figure) of the dispersed spectrum towards the vertically
mounted \ion{Ca}{II} H CCD camera (``blue channel''). The part of the spectrum 
at a longer wavelength (indicated in the figure by a red ray) instead passes below the pick-up mirror towards the channel at 630\,nm (``red channel''). Spectrally undispersed scattered light inside the instrument (indicated in the figure by {\em black arrows}) can reach the 630\,nm
channel CCD only from a small solid angle, whereas most
of the scattered light that hits the pick-up mirror can still reach the
\ion{Ca}{II} H CCD at some place. Because the solar light level in the
near-UV is rather low, especially in the line core of \ion{Ca}{II} H,
additional photons can easily lead to a detector count rate that is
significant with respect to the true solar spectrum. 

The parasitic stray light is generated inside of POLIS. Its light level should therefore scale with the amount of light that enters through the field stop in F2, which can be related to the average of $<I_{\rm local}>$\,. The coefficient $\alpha_3$ in Eq.~(\ref{eq_para}) does not refer to any specific optical element.

Equation (\ref{eq_stray}) can be simplified for any practical
application of a stray-light correction. The full aperture contributing to
$I_{\rm local}(\lambda)$ is rather large, covering more than one supergranular
cell and therefore providing a large-scale average profile similar to a
full-disk spectrum, and there are no means to determine $I_{\rm
  global}(\lambda)$ from the observed spectra. We thus substituted $I_{\rm
  global}(\lambda)$ with $I_{\rm local}(\lambda)$, even if for observations
off the disk center there should be some difference between the two spectra
because $I_{\rm local}(\lambda)$ changes with the telescope pointing. To apply the correction to observations, we here implicitly assume that the average profile  $I_{\rm local}(\lambda)$ is always calculated from quiet Sun regions with a granular pattern. In the case of, e.g., sunspot
observations, this implies that only a subfield of the observed FOV is used
for the average. Because the dark current can be directly measured and
subtracted, the simplified version of Eq.~(\ref{eq_stray}) then reads as
\begin{eqnarray}
I_{\rm obs} (\lambda,x,y) = I_{\rm true} (\lambda,x,y) + \alpha \, I_{\rm  local}(\lambda) + \beta\, <I_{\rm local}>   \label{eq_stray_simp}\,.
\end{eqnarray}
Even if the equation depends solely on two parameters, it turns out that determining $\alpha$ is far from straightforward. In Sect.~\ref{meas_sect}, we describe the results of our several different approaches to determine $\alpha$ and $\beta$.
\subsection{Stray-light correction and spatial deconvolution using a known PSF\label{deconv_sect}}
To apply the stray-light correction, different approaches are
possible. One straightforward method is based on the simplified
Eq.~(\ref{eq_stray_simp}). The knowledge of the instrumental PSF$_{\rm instr}$
also allows, however, more detailed methods to be used that take the spatial dependence of stray light - as given by the second term on the right-hand side of Eq.~(\ref{eq_psf}) - fully into account. For such approaches, we describe a first-order and a full correction by a deconvolution in the following.
\subsubsection{First-order correction} 
To derive a first-order correction for the stray light caused by the instrumental PSF, we reformulate and discretize Eq.~(\ref{eq_img}) as
\begin{eqnarray}
I_{\rm obs}(x,y) = I_{\rm true}(x,y) + \sum_{x^\prime,y^\prime} K(x-x^\prime,y-y^\prime) I_{\rm true}(x^\prime,y^\prime) \nonumber\\
 - \sum_{x^\prime,y^\prime} K(x-x^\prime,y-y^\prime) I_{\rm true}(x,y) \label{win_loss}\\
= (1-\alpha) I_{\rm true}(x,y) + \sum_{x^\prime,y^\prime}
K(x-x^\prime,y-y^\prime) I_{\rm true}(x^\prime,y^\prime)\,, \label{win_loss1}
\end{eqnarray}
where $K$ is the kernel describing the instrumental PSF and the sums are to be executed for all $x^\prime,y^\prime \neq x,y$.

The last two terms on the right-hand side of Eq.~(\ref{win_loss}) describe the
stray light introduced into a location $(x,y)$ from other locations in the
FOV (``gain'' term) and the ``loss'' of intensity to them, respectively. It can be shown that the first-order correction for the gain term is given by 
\begin{eqnarray}
I_{\rm true}(x,y) =  I_{\rm obs}(x,y) - \sum_{x^\prime,y^\prime}
  K(x-x^\prime,y-y^\prime) I_{\rm obs}(x^\prime,y^\prime)  \,,  \label{eq_1stcorr}
\end{eqnarray}
when all terms proportional to $\alpha,\alpha^2, K\,\alpha, K^2$ are neglected
(see Appendix \ref{appa}). Equation (\ref{eq_1stcorr}) thus corresponds to the
approach of calculating the stray light that enters at $(x,y)$ as the observed
intensity in the surroundings weighted by a PSF with a form as in the second term on the right-hand side of Eq.~(\ref{eq_psf}).

For an application of the correction, one has to take into account that for
slit-spectrograph observations the scanning direction (denoted by $x$) and the
direction along the slit (denoted by $y$) are not fully equivalent. All
spectra $I_{\rm obs}(x^\prime,y^\prime)$ for $x^\prime \neq x$ are taken at a
different time $t^\prime$, and only the profiles in the individual
spectrum $I_{\rm obs}(x,y^\prime)$ are obtained simultaneously for all
$y^\prime$. Because the solar scene is changing with time during the scanning,
it can therefore be necessary to restrict the sum in $x^\prime$ to the
temporally ``close'' surroundings instead of using the full range of the known
PSF. The time lag between two scan steps is given by the integration time for
each step $t_{\rm integ}$ times their distance $\Delta x$, where $\Delta
x\cdot t_{\rm integ}$ should be significantly smaller than the typical
granular time scale of about five minutes.

The loss term in Eq.~(\ref{win_loss1}) could in principle be corrected for as
well using the value of $\alpha$, but given the difficulty of achieving high
accuracy in its determination, that is not recommended. If the intensity
normalization of the observed spectra is later done by normalizing an average
profile after the correction to a theoretical or observed reference profile,
the loss term will also be corrected for automatically by the normalization
coefficient. The first-order correction is easy to apply and does not
introduce numerical problems when $K$ is given, despite some eventual
interpolation in the construction of a discrete 2D kernel.  
\subsubsection{Fourier deconvolution} In the Fourier domain, Eq.~(\ref{eq_img}) transforms to
\begin{eqnarray}
\tilde{I}_{\rm obs}  = \tilde{K} \cdot \tilde{I}_{\rm true} \,,
\end{eqnarray}
where\, $\tilde{ }$\, denotes the Fourier transform and only the static
instrumental PSF described by the kernel $K$ is considered. It can be easily
shown \citep[e.g.,][]{deforest+etal2009} that for a known $K$ the reverse direction is given by
\begin{eqnarray}
\tilde{I}_{\rm true} = \tilde{K}^{-1} \cdot \tilde{I}_{\rm obs} \, \label{fft_deconv}
\end{eqnarray}
where $\tilde{K}^{-1}$ is the element-wise inverse operator for the Fourier
transform of $K$. 

The application of Eq.~(\ref{fft_deconv}) exceeds the first-order correction
by taking both the gain and the loss term into account at the same
time. However, even if the derivation of $\tilde{K}^{-1}$ and the subsequent
deconvolution is straightforward, its application can have unwanted side
effects, such as amplifying noise. We compared the Fourier
transform of the inverse PSF for POLIS with the corresponding Fig.~1 of
\citet{deforest+etal2009}. The resulting curve directly resembled the kernel
they obtained {\em after} regularization for the noise suppression. For the
present study we therefore assumed that we do not need to additionally modify
the PSF or apply a high-frequency noise filtering before the deconvolution.
\section{Stray-light measurements\label{meas_sect}}
The following sections describe the measurements for obtaining generic
estimates of the stray-light level and of the instrumental PSF. We also
suggest the best approach to applying the parasitic light correction.
\subsection{Parasitic light \label{para_sect}}
We start with the parasitic light coefficient $\beta$
(Eq.~\ref{eq_stray_simp}) because, unlike $\alpha$, it can be determined
directly. From the discussion of the source of the parasitic light in the
previous section, it follows that if the direct illumination of the CCD with
sunlight is blocked, only the randomly scattered light will be left over. This
is thus the method of choice in determining $\beta$. We therefore compared measurements of a spectrum with a fully open aperture and a spectrum with the light path blocked close to the camera mirror inside of POLIS (Fig.~\ref{fig10}). The location of the blocking prevented the direct illumination of the CCDs with the resolved spectrum, but did not affect the stray-light level inside the spectrograph otherwise. Both measurements were corrected for the dark current before the analysis. 

To quantify the amount of parasitic light as a fraction of some mean intensity
of the incoming light, we averaged the intensity of the full-aperture spectrum
over a pseudo-continuum window in the spectrum from 396.383\,nm to 396.393\,nm
as reference, and this yielded a mean value of 2716 counts. The average
intensity value of the measurement without direct illumination of the CCD was
around 148 detector counts. The corresponding image was homogeneously lit and
showed no trace of spectral lines which ensures that no dispersed stray light
was included in the measurement. The ratio of parasitic and reference intensity
then is about 5\,\% in the \ion{Ca}{II} H channel. The corresponding values
for the channel at 630\,nm were 20380 and 58 counts, respectively, yielding a
fraction of 0.3\,\% that is presumably negligible in an analysis of the 630\,nm data. 

For all following stray-light measurements in the \ion{Ca}{II} H channel, we have corrected the spectra for the parasitic light separately spectrum by spectrum. We determined the average intensity in the continuum window in each spectrum, averaging along the slit as well, and then subtracted 5\% of this average value in detector counts from all intensity values of the spectrum. 
\begin{figure}
\centerline{\resizebox{8.8cm}{!}{\includegraphics{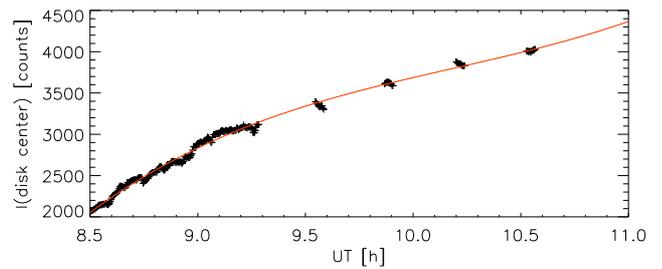}}}
\caption{Intensity normalization curve for the observations of 14/08/2009. The observed intensity at disk center is given by the {\em black crosses}, the {\em red line} is a polynomial fit to the data points.\label{fig8a}}
\end{figure}

For the correction of scientific observations, we suggest using a slightly modified approach. From repeated small maps taken with about half an hour cadence at
the center of the solar disk, one can obtain an intensity normalization curve
for each point of time during the day. Figure \ref{fig8a} shows an example of
such a curve for 14 August 2009. The first observation of this day
was a large-area scan on disk center, so its intensity was used
for the calibration curve, too. For each separate scan of a certain FOV,
one then calculates the average profile over all scan steps, first dividing
each spectrum by the corresponding value of the normalization curve. This
removes the temporal trend during the scan and provides an average spectrum
$I_{\rm local}$ with an intensity only proportional to that of the observed
FOV. The parasitic correction is then given by the 5\%-fraction of the
intensity in the continuum window of the average profile. To include the
change in light level during the observation again -- because $<I_{\rm
  local}>$ at a given moment of time scales with it --  the normalization
curve from the start until the end of the scan is normalized to its
mean value inside that time span. This provides the relative variation in the light level, hence of the parasitic correction for a spectrum taken at a
given moment inside the FOV that entered into the calculation of $I_{\rm
  local}$. Multiplying the parasitic correction with the relative temporal
variation in intensity during the scan yields a correction for the
parasitic stray light that accounts for both the variation in $I_{\rm
  local}$ with the telescope pointing and the temporal variation in the light level during the day. 
\subsection{Spider mounting\label{spider}}
\begin{figure}
\centerline{\resizebox{8.8cm}{!}{\includegraphics{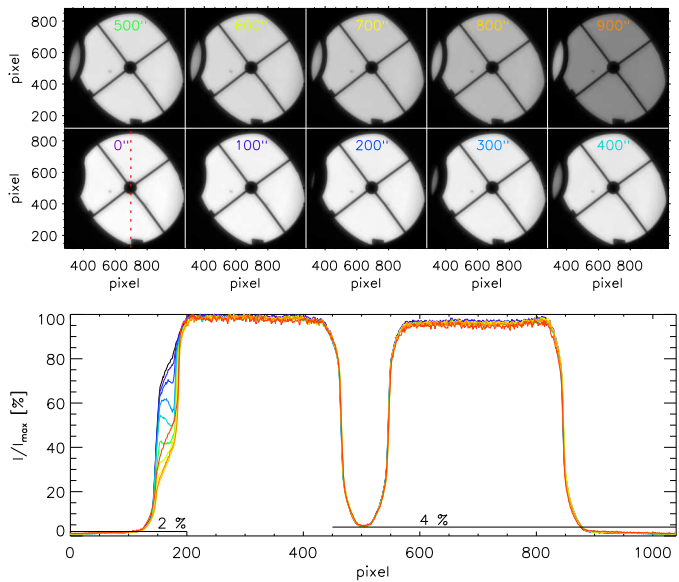}}}
\caption{Images of the pupil plane taken with the PCO setup in a CLV run at 557\,nm. {\em Top panel}: pupil images, all shown in an identical display range. The offset of the telescope pointing from the disk center is given in each subpanel in arcsec. {\em Bottom}: intensities of each pupil image along the cut marked by the {\em red vertical} line in the image at 0$^{\prime\prime}$.\label{fig3}}
\end{figure}

The guiding telescope of the VTT is fed by a small pick-up mirror located
close to the entrance window of the evacuated telescope tube. The pick-up
mirror is fixed to a spider mounting and produces a central obscuration in any
pupil plane at the VTT. We set up an imaging channel to determine the light level inside the central obscuration. We deflected the light with the 45-deg mirror to an optical rail where we mounted a lens for creating a pupil image, a small-band interference filter at 557.6\,nm, and some neutral
density (ND) filters to reduce the intensity to a suitable level for the PCO Sensicam camera used. With this setup, we took images of the spider mounting in the
pupil plane for a CLV run similar to the scientific observations, to measure
the stray-light level and at the same time to identify a possible variation with the telescope
pointing. The {\em upper panel} of Fig.~\ref{fig3} shows the
resulting pupil images when moving across the solar disk in steps of
100$^{\prime\prime}$. We took cuts through the images to determine the
residual intensity in the shaded areas and normalized each of the cuts to its
maximum intensity value  ({\em lower panel}). The pupil images show a change
in the global light level with the telescope pointing, but the intensity
curves after the normalization are nearly identical. The residual
intensity in the central obscuration is always about 4\%. The light level at the border of
the FOV (``tail''), which can only be caused by stray light, is below 2\%. These numbers now, however, refer to a pupil plane, not a focal plane. After the
reflection on the 45-deg mirror, the light passed through only three optical
elements (lens, interference filter, ND filter). Because the spider mounting is located
behind the entrance window,  the stray light caused by the two coelostat mirrors and the entrance window is absent.
\subsection{Methods without explicit measurements \label{wo_meas}}
There are three more methods that provide an estimate of stray light without
requiring explicit measurements. The first is to match average observed
spectra to an atlas profile, as suggested by \citet{allendeprietoe+etal2004}
and \citet{cabrerasolana+etal2007}. A comparison with the presumably
stray-light-free spectra of the FTS spectral atlas
\citep{kurucz+etal1984,brault+neckel987} yielded the best match for a value of
$\beta= 20\,\%$ for both channels of POLIS. The approach is, however, not
fully consistent with Eq.~(\ref{eq_stray_simp}) because it implies that all
stray light is of parasitic nature. Instead, the direct measurement of $\beta$
of Sect.~\ref{para_sect} excludes the high value retrieved with this
approach. The approach is also aimed at determining the {\em spectral} point
spread function of the optics behind the slit, because it uses an average observed profile without spatial resolution and the comparison of its shape to a reference profile. It therefore measures all spectral resolution effects of the
spectrograph in addition to the parasitic contribution.

A second method is to compare profiles from the umbra of sunspots with only
a little intrinsic radiation to an average quiet Sun
spectrum. \citet{reza+etal2007} determined a level of at maximum 12\,\% of
stray light at 396\,nm for the sum of $\alpha+\beta$ in the umbra, which can
be taken as the minimum amount of stray light in the brighter quiet Sun
surroundings. A simultaneous inversion of 630\,nm and 1565\,nm spectra with the SIR code yielded a value of $\alpha=$\,10\% of stray light in the umbra \citep{beckthesis2006,beck2008}. 

The last indirect method is the presence of residual polarization signal in
the telluric line blends of the 630\,nm channel. In the data reduction of
long-integrated ($>20$ secs) spectra of the 630\,nm channel, the telluric
lines at 630.20\,nm and 630.27\,nm sometimes showed a non-zero polarization
signal even if the nearby continuum wavelengths showed none. This puzzling
behavior could be traced back to the $I\rightarrow QUV$ cross-talk correction
in combination with the real-time demodulation of the Stokes vector in
POLIS. This demodulation uses pairs of image {\em differences} for Stokes
$QUV$, but the addition of all images for Stokes $I$. Parasitic light thus
adds up in $I$ but cancels in $QUV$. For a correction of the cross-talk, one
then has to use a reduced intensity $I_{\rm true} = I_{\rm obs}-{\rm const.}$
in the calculation of the  cross-talk coefficient $c_{I\rightarrow QUV} =
QUV_{\rm obs}/ I_{\rm true}$ at continuum wavelengths. Subsequently, also
$c_{I\rightarrow QUV}\cdot I_{\rm true}$ has to be subtracted from $QUV$. This
additional correction was used in the data reduction of POLIS data in 2003
with const.\,$\equiv \beta =\,4$\,\% of $I_c$, but only for the
long-integrated data. For shorter integration times, the effect was below the
noise level and could not be detected. Because the stray-light covers in POLIS
were modified afterwards as a consequence of this finding, the number can only
serve as an order of magnitude estimate. We point out that this effect produces spurious polarization signals inside spectral lines caused by cross-talk from Stokes $I$ {\em even when} close-by continuum wavelengths are forced to zero polarization.
\begin{figure}
\centerline{\resizebox{8.8cm}{!}{\includegraphics{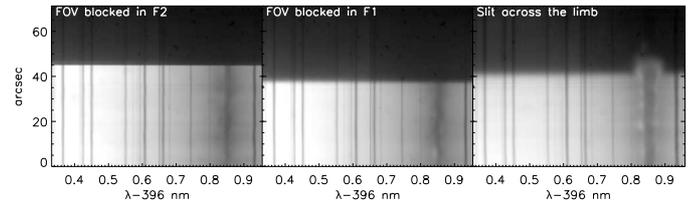}}}
\caption{POLIS \ion{Ca}{II} H spectra with partly illuminated FOV, logarithmic display range. {\em Left}: FOV blocked in F2 (hairline mounting). {\em Middle}: FOV blocked in F1 (filter wheel). {\em Right}: slit across the solar limb.\label{fig4}}
\end{figure}
\subsection{FOV partially blocked in the focal plane (spatial PSF)\label{sect_psf}}
\begin{figure}
\centerline{\resizebox{8.8cm}{!}{\includegraphics{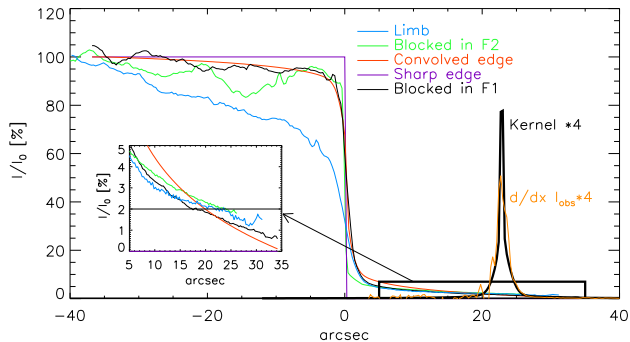}}}
\caption{Observed intensity variation along the slit for FOV blocked in F1
  ({\em black}), blocked in F2 ({\em green}), and for the slit across the limb
  ({\em blue}). The {\em purple lines} at $x,y$ = (0$^{\prime\prime},100\%$)
  denote a simulated sharp edge that after convolution with the kernel at the
  {\em right} yields the {\em red curve}. The inlet {\em on the left} shows
  a magnification of the stray-light tail region. The derivative of the
  observed intensity variation for the FOV blocked in F1 is overplotted over
  the kernel as an {\em orange line}.\label{fig5}}
\end{figure}

To derive the spatial PSF of POLIS, we took spectra with POLIS where part of the FOV
in the focal plane was blocked by a metal plate. In addition to blocking part
of the FOV in the focal planes F1 (telescope focus) and F2 (entrance of POLIS), we also took spectra with the slit placed across the
solar limb. Figure \ref{fig4} shows example spectra of the three cases. The
blocking edge is sharper for the blocking in F2 inside of POLIS. This is
caused by fewer optical elements being located behind F2, by the
impossibility of using the adaptive optics while blocking in F1 and by the
fact that the metal plate had to be placed at some distance to F1 ($\approx 1$
cm). All spectra are displayed on a logarithmic scale, otherwise the stray light could not be seen at all directly in the images. We then averaged the intensity in the same continuum window near 396.4\,nm as before to obtain the spatial intensity variation across the blocking edge and the limb, respectively (Fig.~\ref{fig5}). 

To determine the spatial PSF, we then defined an artificial sharp edge  located appropriately along the slit, and constructed a model
for the spatial PSF from a combination of a Gaussian and a Lorentzian
curve. We convolved the sharp-edge function with the kernel and modified the
free parameters of the two functions used in its generation by trial-and-error
until the convolved edge function ({\em red line}) matched the observed
intensity trend across the edge for the blocking in F1. The finally used
kernel is shown in the {\em  right} part of Fig.~\ref{fig5} at about
$x=+25^{\prime\prime}$; it drops to basically zero at about
10$^{\prime\prime}$ distance from its center. The area of the kernel is
normalized to unity, and its maximum value then is around 20\%. For better visibility, it is shown multiplied by four in the figure. The
shape of the kernel can also be cross-checked against the observed intensity
variation by another method than the convolution of the sharp-edge
function. In case of an axisymmetric 2D kernel $K(x,y)$, it can be shown that the 1D kernel $K(x)$ is given by
\begin{equation} 
K(x) = \frac{{\rm d} I_{\rm obs}}{\rm dx}(x) \,,
\end{equation}
when the true intensity corresponds to a step function and $I_{\rm obs}(x)$ is the observed intensity variation across the step function \citep{collados+vazquez1987,bonet+etal1995}.

We overplotted the derivative of the observed intensity variation for the
blocking in F1 over the kernel in the right part of Fig.~\ref{fig5}. The match
is reasonably close, with a slightly lower central value and a broader central core in the derivative. The trial-and-error method for defining the kernel thus yielded a good match to the theoretical expectation. 
 
Since the chosen kernel exhibits a sharp drop, most of the local stray
light should come from the close vicinity  ($d<2^{\prime\prime}$) of a given
point $(x,y)$. The inlet in the {\em left-hand} part of Fig.~\ref{fig5} shows
that, actually,  the tail level of the observations is not matched well, because the curve of the convolved sharp edge reaches zero several arc
seconds before the observations. The residual intensity in the far tail
($>20^{\prime\prime}$) is at maximum 2\%. We point out that the observations
of the solar limb and with the FOV blocked in F1 only differ slightly for
$x>0^{\prime\prime}$, which implies that most of the stray light is caused by
the instrumental PSF and not the telescope or atmospheric scattering. 

For the application of Eqs.~(\ref{eq_1stcorr}) and (\ref{fft_deconv}), a 2D kernel is needed. We used the assumption of axial symmetry to create the
2D kernel $K$ from the 1D version shown in Fig.~\ref{fig5}, and
divided $K(x,y)$ by $2\,\pi\,r$, with $r=\sqrt{x^2+y^2}$, to maintain the
normalization of the area of the 1D kernel.
\subsection{Summary of stray-light measurements} 
We used different methods to estimate the amount of stray light in POLIS. The
upper two sections of Table \ref{tab1} list the stray-light values determined
in this work. In the lower section, we added the values that have been
presented in previous studies for any of the two channels of POLIS. The
definition of the stray-light fraction differs in most cases, but the umbral
profiles provide a {\em lower} limit of 10\,--\,12\% for the total stray-light
contamination in quiet Sun areas without strong spatial intensity
gradients. The parasitic light fraction could be accurately determined as 5\,\% (0.3\,\%) of the intensity for the channel at 396\,nm (630\, nm)
for the data taken in 2009, but it presumably was higher in previous
years before the modification of the stray-light covers inside the
spectrograph. The stray-light level at distances above $20^{\prime\prime}$
from any lit area is on the order of 2\% only, and the spatial PSF is strongly peaked with significant contributions only up to about $2^{\prime\prime}$. There were no indications that the relative stray light in the telescope changes somehow with the telescope pointing in a CLV run. The tail of the observed intensity variation across a sharp edge can be used as a proxy for the reduction of the stray-light level in off-limb spectra. 
\begin{table}
\caption{Stray-light levels for the VTT/POLIS in various studies. \label{tab1}}
\begin{tabular}{c| cc}
Measurement & Wavelength & Stray-light level  \cr \hline\hline
Pupil plane (PCO)& 557\,nm& 4 \% ($\sim$ 1-2 \% tail) \cr\hline\hline
\multicolumn{3}{c}{POLIS (this work)}  \cr \hline
F1 blocked& 396\,nm& $\sim$ 1-2 \% (tail)  \cr
F2 blocked& 396\,nm& $\sim$ 1-2 \%  (tail)\cr
Limb & 396\,nm& $\sim$ 1-2 \%  (tail)\cr
Blocked behind grating& 396\,nm & $\sim$ 5 \%  \cr
Blocked behind grating& 630\,nm & $\ll$ 1 \% \cr
Convolution of FTS& 630/396\,nm& 20 \%  \cr\hline
\multicolumn{3}{c}{POLIS (other studies)}  \cr \hline
Umbral profile \tablefootmark{1}& 396\,nm & $\sim$ 12 \%  \cr
Room light \tablefootmark{1} & 396\,nm & $<$ 1 \% \cr
Convolution of FTS \tablefootmark{2} & 630\,nm & 15 \%  \cr
Umbral profile \tablefootmark{3} & 630\,nm & $\sim$ 10 \%  \cr
Parasitic light correction \tablefootmark{4} & 630\,nm & $\sim$ 4 \%  \cr\hline
\end{tabular}\\$ $\\
\tablefoottext{1}{\citet{reza+etal2007}} \;\; \tablefoottext{2}{\citet{cabrerasolana+etal2007}}\;\; \tablefoottext{3}{\citet{beck2008}}\;\; \tablefoottext{4}{Keyword in POLIS data reduction software, used for long-integrated data in 2003.}
\end{table}
\section{Application to data\label{deconv}}
As a first application to observed data, we compare the performance of the
stray-light corrections with and without explicitly using the instrumental PSF
on a data set taken near the solar limb. The application to the limb data set actually yielded improved values for the stray-light contribution $\alpha$ because the residual intensity beyond the limb provides a good reference for the quality of the correction. In the second and third examples, we instead test the effects of the deconvolution as a method to improve the spatial resolution of observations on spectroscopic and spectropolarimetric data, respectively.
\subsection{Example A: application to a limb data set}
To test both the stray-light correction and the spatial deconvolution, we
used a data set that was taken near the solar limb on 25 August 2009,
with an FOV that extended beyond the limb (see Fig.~\ref{appl_2d} below). This offers a good opportunity to determine the quality of the
corrections because the limb is similar to a sharp edge and beyond the
limb the (pseudo-)continuum intensity should be zero. The map was
taken with 134 steps of 0\farcs3 step width and an integration time of 13
seconds per scan step \citep[for more details see][]{beck+etal2011}. We applied a variation of corrections to the previously flat-fielded data to assess the effects of each step on the data. We used the following combinations of corrections:
\begin{itemize}
\item[1.] Parasitic light only ($\beta$).
\item[2.] Parasitic light ($\beta$) with subsequent first-order correction (Eq.~\ref{eq_1stcorr}).
\item[3.] Parasitic light ($\beta$) with subsequent Fourier deconvolution (Eq.~\ref{fft_deconv}).
\item[4.] Parasitic light ($\beta$) and stray light ($\alpha$) as given by Eq.~(\ref{eq_stray_simp}) (more details below).
\end{itemize}
Some of the approaches had to be modified slightly for treating off-limb
spectra. Method no.~1 serves as a reference of basically uncorrected
spectra. For the first-order correction in Method no.~2, we restricted
  the sum of Eq.~(\ref{eq_1stcorr}) to $(x^\prime,y^\prime) = (x,y) \pm
  (5,40)$ pixels to avoid using scan steps that were taken at a significantly
  different time. The Fourier deconvolution in the 3$^{rd}$ method was done according to Eq.~(\ref{fft_deconv}) by multiplying the Fourier transform of the 2D maps of the intensity at each wavelength, $I(\lambda)$, with the inverse of the kernel.
\begin{figure}
\centerline{\resizebox{8.8cm}{!}{\includegraphics{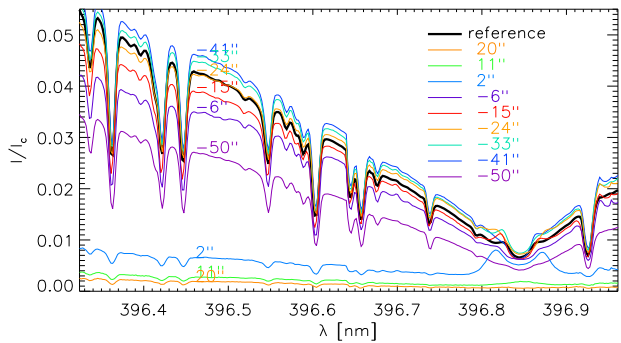}}}
\caption{Comparison of a large-scale average profile multiplied by $\alpha$ ({\em thick black}) and individual local stray-light profiles calculated around a given pixel using the PSF. The labels denote the distance of the pixel to the limb in arcsec (negative/positive $\equiv$ on/off-disk). \label{stray_comp}}
\end{figure}
\begin{figure}
\centerline{\resizebox{8.8cm}{!}{\includegraphics{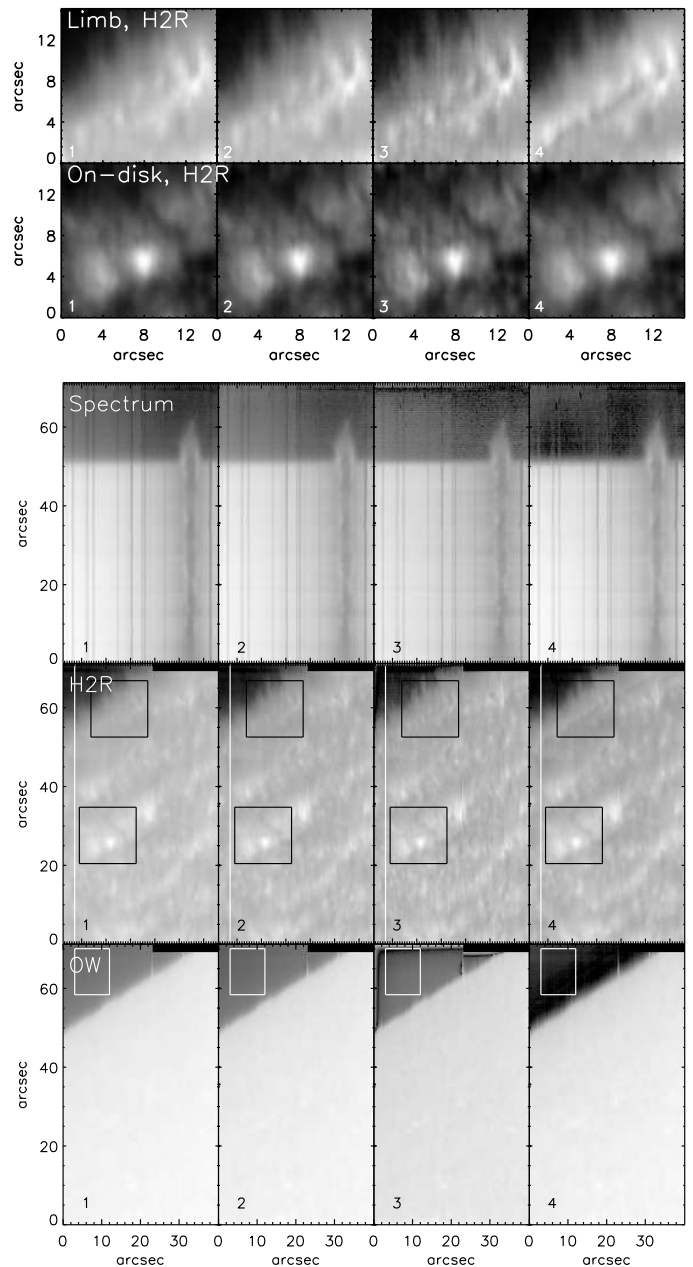}}}
\caption{2D maps and spectra after the correction. {\em Lower panel:} OW map in logarithmic display ({\em bottom row}), H$_{\rm 2R}$ map ({\em middle row}), spectra ({\em top row}) along the cuts marked by {\em white vertical lines} in the H$_{\rm 2R}$ map. {\em Upper panel}: magnification of the areas marked by {\em black squares} in the H$_{\rm 2R}$ map (on-disk and at the limb). The {\em white rectangles} denote the off-limb area used to derive the profiles of Fig.~\ref{offlimb_spec}.\label{appl_2d}}
\end{figure}

For Method no.~4, we used a constant value of $\alpha = 10\%$ as the minimum level of stray light for all spectra on the disk and an
average local profile $I_{\rm  local}$ calculated from the part of the FOV
that was most remote from the limb. The intensity beyond the limb drops fast, however, with the limb distance (Fig.~\ref{fig5}), so subtraction of a fixed
$\alpha\,I_{\rm  local}$ yielded negative intensity values beyond the limb. To avoid this overcompensation, we scaled $\alpha$ down with
increasing limb distance using the variation in intensity for the
observation with the FOV blocked in F1 (Fig.~\ref{fig5}). We took the observed intensity variation beyond the location
of the edge ($x<0^{\prime\prime}$) and normalized the first point to 10\,\%
to have a smooth connection to the correction on the disk. The limb location was set to where the intensity dropped below 5\,\%. 

It turned out that a good off-limb correction was only achievable when the
intensity of $I_{\rm  local}$ was increased by a factor of two. Because
  the correction uses $\alpha\,I_{\rm  local}$, the factor of two can be
  attributed either to $\alpha$ {\em or} $I_{\rm  local}$. Using an average
disk center profile with its intrinsically higher intensity and keeping
  $\alpha$ at 10\,\% yielded a worse result, because the spectral line blends
in the wing were located at slightly different wavelengths than in the
observed FOV, and it overcompensated for the stray light present as
well. We therefore used an on-disk correction corresponding to $\alpha\,I_{\rm
    local}$ with $\alpha = 20\,\%$, with the off-limb correction normalized to
  20\,\% on the first point.  

The values used in the correction can actually be cross-checked with the
PSF. Figure \ref{stray_comp} shows $\alpha\,I_{\rm  local}$ used in Method
no.~4 on the disk, together with some profiles
corresponding to the gain term of Eq.~(\ref{eq_1stcorr}), calculated locally
from the surroundings of each pixel in the FOV weighted with the PSF. The
stray-light contributions predicted by the gain term match the generic $\alpha\,I_{\rm local}$ well, as long as the pixel, whose surroundings were used, was more remote from the limb than 10$^{\prime\prime}$. The similarity of the stray-light profiles calculated using the PSF to each other and to the average profile supports using a single average profile $I_{\rm  local}$ in Eq.~(\ref{eq_psf}) ignoring the PSF, but with a doubled $\alpha$ of 20\,\%. The profile at 50$^{\prime\prime}$ limb distance has a lower intensity because it is at the lower border of the FOV, where the full area of the kernel could not be used.

Figure \ref{appl_2d} shows the result of applying all four methods (each
shown in a column of the figure, in the same order as presented in the text)
to the data set in 2D maps of the FOV and one sample spectrum. We derived one
map in a pseudo-continuum window in the outer wing \citep[OW, see,
e.g.,][]{beck+etal2008} of the \ion{Ca}{II} H line and one map averaged over
the wavelength region of the $H_{\rm 2R}$ emission peak. The amount of
residual off-limb stray light can be clearly seen in the OW map ({\em bottom
  row}) and the spectra ({\em third row}). Only Method no.~4 removes the off-limb stray light satisfactorily.
 \begin{figure}
\centerline{\resizebox{8.8cm}{!}{\includegraphics{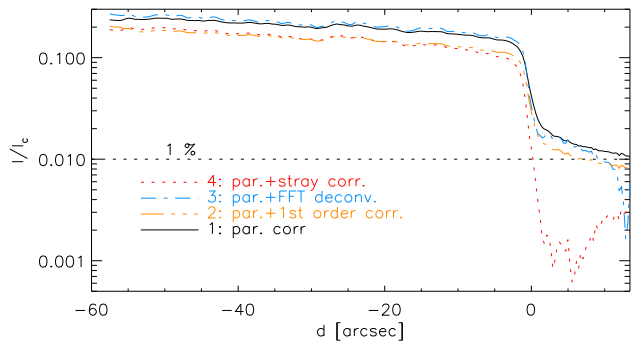}}}
\caption{Intensities across the limb in the OW wavelength range for an
    individual spectrum. {\em Solid black line}: Method no.~1. {\em
        Triple-dot-dashed orange line}: Method no.~2. {\em Dash-dotted
        blue line}: Method no.~3. {\em Dotted red line}: Method no.~4.\label{across}}
\end{figure}
\begin{figure}
\centerline{\resizebox{8.8cm}{!}{\includegraphics{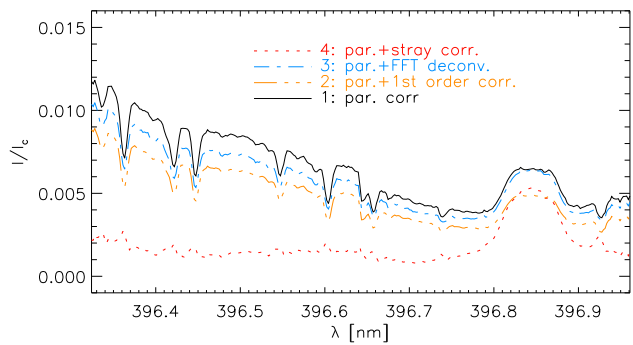}}}
\caption{Average off-limb spectra from the area marked by a {\em white
    rectangle} in Fig.~\ref{appl_2d}. Line styles and colors are as in
    Fig.~\ref{across}. \label{offlimb_spec}}
\end{figure}

The residual stray-light level beyond the limb is quantified in
Fig.~\ref{across}, which shows the intensity variation along the slit in the
OW map for a randomly chosen scan step. The on-disk light level is globally
reduced for Methods no.~2 and 4, whereas the Fourier deconvolution
(no.~3) keeps the same level as the uncorrected spectra (no.~1). This reflects
that only with the Fourier deconvolution is the loss term of
Eq.~(\ref{eq_1stcorr}) taken into account, whereas all other methods only
correct the gain term. Beyond the limb, the stray-light level stays at the
value of the uncorrected spectra of about 1\,--\,2\,\% for the Fourier
deconvolution (no.~3), whereas the first-order correction (no.~2) and the
ad-hoc stray-light correction (no.~4) reduced the off-limb light level. The last method yields a correction an order of magnitude better than all others, reaching a level of about 0.2\,\% of residual intensity.

The average profiles shown in Fig.~\ref{offlimb_spec}, calculated for the off-limb area marked in the OW map of Fig.~\ref{appl_2d}, also indicate the quality of the correction. These profiles allow instant differentiation of
 under/overcorrected stray light by the shape of the line blends
in the wing:  absorption profiles indicate residual stray light, whereas
emission profiles for the photospheric blends at these heights above the limb
indicate an overcorrection. Only Method no.~4 yields fully negligible
residuals of the line blends. For the three other methods, absorption lines
are seen in the wing. For the emission in the \ion{Ca}{II} H line core --
basically, the only scientifically interesting quantity beyond the limb -- Methods no.~2 and 4 yield a similar amplitude, whereas the Fourier deconvolution (no.~3) yields nearly the same intensity as the uncorrected spectra (no.~1).

Even if Method no.~4 gives the best results for the off-limb stray-light
correction, this does not signify that the principle of the other
methods is wrong. Methods no.~2 and 3 follow the initial way in which stray light is created closer than Method no.~4. Both approaches, however, only correct the static contribution to the stray light caused by the instrumental
PSF, within the limits of accuracy in its derivation. Their worse performance
beyond the solar limb implies that in this case both the residual dynamical
part of the PSF beyond the AO correction and the PSF of the telescope and the
optics upfront of F1 play a role. Because the true solar off-limb
intensity outside of the very emission core is zero, any residual stray light
not accounted for by the instrumental PSF is seen prominently in
  Figs.~\ref{appl_2d} or \ref{offlimb_spec}. The ad-hoc method (no.~4) instead seems to correct well for both the stray light caused by the instrumental PSF, as well as for the additional stray-light contributions, but only with a static correction without taking fluctuations of the atmospheric scattering into account.
\begin{figure}
\centerline{\resizebox{8.8cm}{!}{\includegraphics{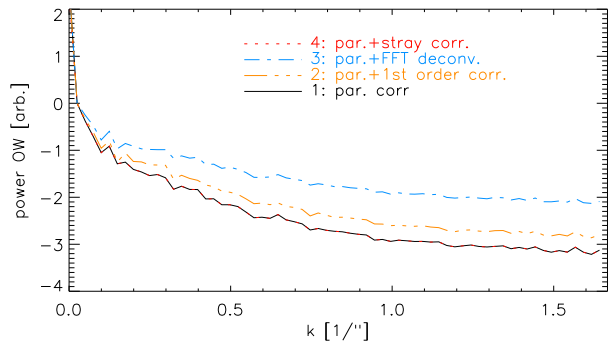}}}
\caption{Fourier power spectrum of the OW map vs.~spatial frequency. Line
    styles and colors are as in Fig.~\ref{across}.\label{spat_res}}
\end{figure}
\begin{table}
\caption{Spatial rms contrast and spectral rms noise in \% of $I_c$.\label{tab_rms}}
\centering
\begin{tabular}{c|cccc}\hline\hline
method   & 1   & 2    & 3    & 4 \cr\hline
contrast & 5.2 &  5.8 & 6.7 &  6.5  \cr
noise   & 1.15 &  1.18 & 1.34 & 1.17 \cr\hline
\end{tabular}
\end{table}
\begin{figure*}
\sidecaption
\resizebox{11.8cm}{!}{\includegraphics{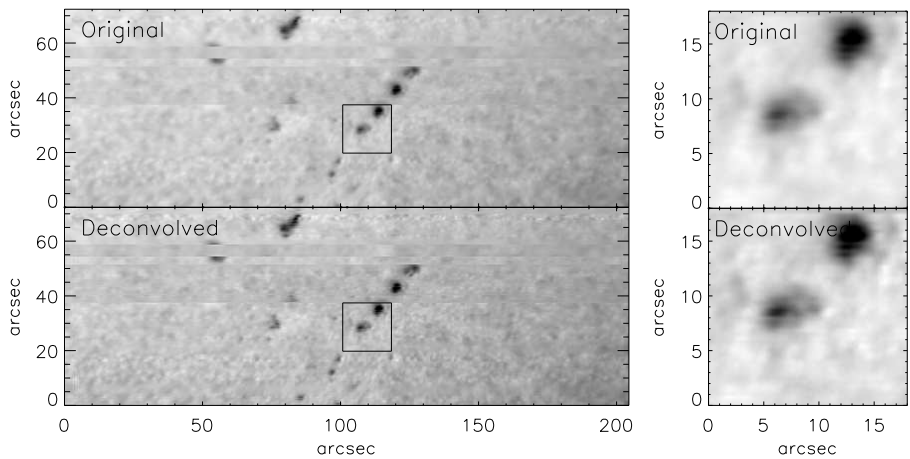}}
\caption{Original ({\em top row}) and deconvolved ({\em bottom row}) continuum intensity maps at 777\,nm of an active region with several pores. {\em Left}: full FOV. {\em Right}: magnification of the region marked by a {\em black rectangle}. \label{777_fov}}
\end{figure*}

As far as the spatial resolution is concerned, both the first-order correction
(no.~2) and the Fourier deconvolution (no.~3) have led to an
improvement, even if it is not very noticeable ({\em second} and {\em third columns} in the {\em top two rows} of Fig.~\ref{appl_2d}). The
Fourier deconvolution had a stronger effect on the visual impression. This is verified by the plot of the spatial Fourier power for the OW map shown in
Fig.~\ref{spat_res}. Only the lower part of the OW map up to $y=
  40^{\prime\prime}$ was used. The power at higher spatial frequencies is
  slightly enhanced after the application of the first-order correction  and
  significantly enhanced after the Fourier deconvolution. The improved spatial
  resolution is reflected also in the root-mean-square (rms) contrast of the
  OW maps (Table \ref{tab_rms}). This number, which measures the relative
  intensity variation, has, however, to be taken with caution because any correction that reduces globally the intensity automatically increases the rms contrast. We also determined the rms variation inside wavelength windows of about 10\,pm extent in the OW spectral region and found it to stay basically at about 1.2\,\% of $I_c$, with an average value of the intensity itself of about 20\,\% at these wavelengths. The Fourier deconvolution (no.~3) has slightly increased the noise.
\begin{figure*}
\sidecaption
\resizebox{11.8cm}{!}{\includegraphics{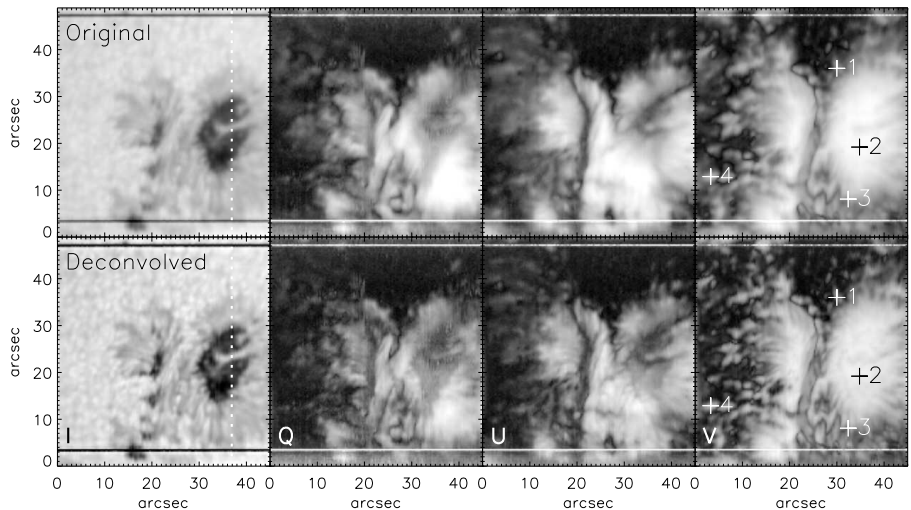}}
\caption{2D maps of the sunspot observations at 630\,nm. {\em Top/bottom row}: original/deconvolved spectra. {\em Left to right}: Stokes $IQUV$. The {\em white dashed line} marks the location of the cut shown in Fig.~\ref{spot_cut}; the {\em crosses} and the corresponding numbers label the locations of the profiles shown in Fig.~\ref{spot_spec}. The horizontal black/white lines at the upper and lower border of the FOV are caused by the hairlines of POLIS.\label{spot_2d}}
\end{figure*}
\subsection{Example B: application to pore observations at 777\,nm}
In another observing campaign in November 2009, we took identical
  measurements with a partly blocked FOV with the main spectrograph of the VTT in a wavelength range around the \ion{O}{i} triplet at 777\,nm. The corresponding convolution kernel derived from the measurement is shown in Appendix \ref{appb}. The spatial sampling along the 0\farcs18 wide slit was 0\farcs18 in this case.

We then applied the spatial deconvolution to an observation of an active
region with several pores, using Eq.~(\ref{fft_deconv}) without any additional
stray-light correction. The spatial scanning in the observation was done with
201 steps of 0\farcs36 width, undersampling the slit width. Figure
\ref{777_fov} shows the full FOV before ({\em upper left panel}) and after the
deconvolution ({\em lower left panel}). The seeing was moderate to bad during
the scan, producing frequent failures of the AO system, with jumps of the FOV
as well. The magnification of two of the pores ({\em right panels}),
however, shows that the deconvolution has not only increased the contrast by a reduction of the intensity, but also leads to an improvement in the sharpness of the structures, e.g., for the boundaries of the darkest parts of the pores.  

\subsection{Example C: application to polarimetric observations}
To test the effect of the deconvolution by Eq.~(\ref{fft_deconv}) on
polarimetric data, we applied it to observations taken with POLIS in its
red channel at 630\,nm. Assuming that the PSF derived for the blue channel is
characteristic of the instrument, we did not determine a new PSF but
used the same one as before. Appendix \ref{appb} shows that applying
this kernel to a knife-edge observation at 630\,nm also provides a
satisfactory match. The observation we discuss here consists of a scan with
150 scan steps of 0\farcs3 step width across an active region
(Fig.~\ref{spot_2d}) done on 6 July 2005, with AO correction. The
integration time was ten seconds per scan step. In this case, the Fourier deconvolution was applied separately to each wavelength and Stokes parameter. 

\begin{figure}
\centerline{\resizebox{8.8cm}{!}{\includegraphics{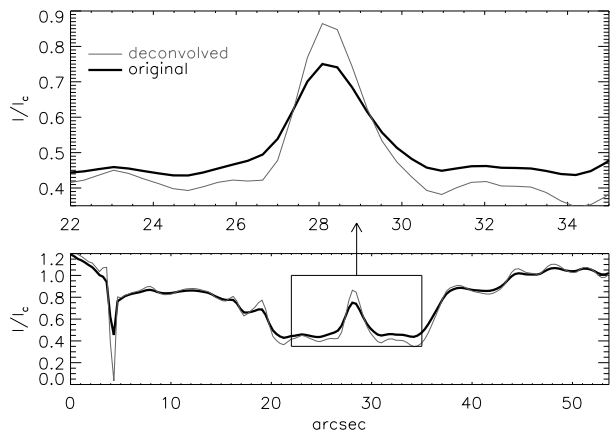}}}
\caption{Cut through the intensity maps of Fig.~\ref{spot_2d}. {\em Thick black}: original data. {\em Thin grey}: deconvolved data. The {\em upper panel} shows a magnification of the central region where the cut crossed a light bridge. \label{spot_cut}}
\end{figure}

In the visual impression of Fig.~\ref{spot_2d}, the improvement in the spatial
resolution is prominent, both in the continuum intensity map ({\em
  left}) and the maps of the absolute wavelength-integrated polarization
signal, $\int |QUV|/I d\lambda$. The umbral dots inside the umbra of the
bigger spot can be seen well in the deconvolved data set and the filamentary
structure of the penumbra of the smaller spot is much clearer. The rms contrast in the continuum image increased from 3.7\,\% to 5.3\,\%. Figure \ref{spot_cut} demonstrates that the increase in the rms contrast is not only caused by the removal of the stray-light contamination. It shows a cut through the umbra of the larger sunspot where a light bridge was located. The full width at half maximum of the light bridge is slightly reduced and the intensity values to the left ($x\sim\,26^{\prime\prime}$) and right  ($x\sim\,30.5^{\prime\prime}$) of the central region decrease to the benefit of a more concentrated distribution with a stronger intensity peak in the deconvolved data, compared to the original data. 
\begin{figure}
\centerline{\resizebox{8.8cm}{!}{\includegraphics{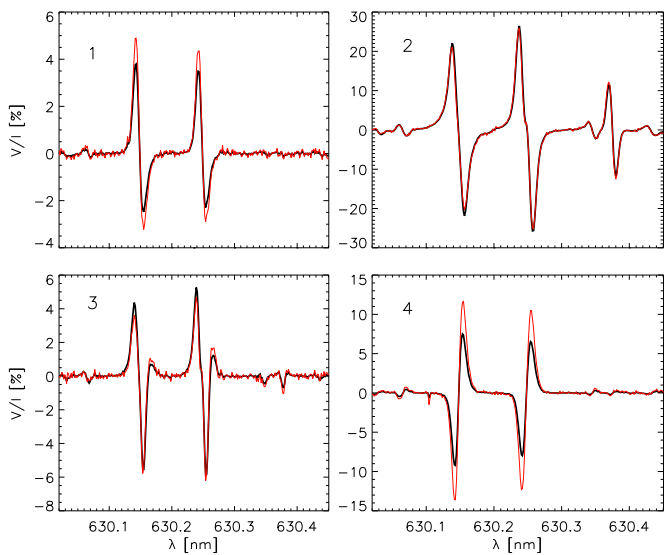}}}
\caption{Stokes $V$ spectra from the sunspot observation. {\em Red}:
  deconvolved spectra. {\em Black}: original spectra. The numbers in {\em the
    upper left} of each panel refer to the labels in Fig.~\ref{spot_2d} that indicate the locations of the spectra in the 2D maps of the sunspot observation. \label{spot_spec}}
\end{figure}

The polarization signal was also usually amplified by the deconvolution. Figure \ref{spot_spec} shows four Stokes $V$ profiles before and after the deconvolution; taken from the locations marked in Fig.~\ref{spot_2d}. On average, the Stokes $V$ amplitudes increased relatively by about 5\,\% of their previous value. These spectra, however, also clearly show the drawback of the Fourier deconvolution: the noise level is amplified as well. In Stokes $I$, the rms noise in a continuum window increased from 0.05\,\% to 0.07\,\% at an average value of about 15.\,000 counts. In Stokes $V$, where the noise level is more critical in the analysis of the data, the rms noise was 0.05\,\% of $I_c$ before and 0.12\,\% of $I_c$ after the deconvolution, respectively. The rms noise in Stokes $V$ is still acceptable, even after the increase by a factor of 2.4. For comparison, \citet{steiner+etal2010} found an increase of the rms noise from 0.1\,\% to 0.2\,--\,0.3\,\% in the deconvolution of IMaX data. 
\section{The relation between PSF and (local) stray  light \label{alpha_true}}
\begin{figure}
\centerline{\resizebox{8.8cm}{!}{\includegraphics{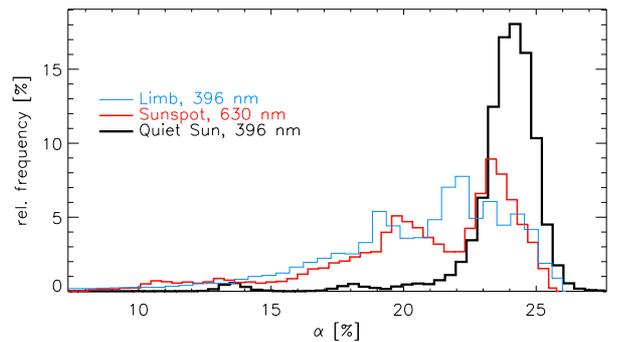}}}
\caption{Histograms of the local value of $\alpha$ in a QS map ({\em thick black}), a sunspot map ({\em red}), and the limb data set ({\em thin blue}).\label{hist_alpha}}
\end{figure}

In the stray-light correction of the off-limb spectra, it turned out by
trial-and-error that a value of $\alpha=10$\,\% is definitely too low. The
finally used stray-light level that yielded a good correction corresponds to
using $\alpha=20$\,\%. The integration of the local stray-light contribution
using the PSF  finally returned the same level (Fig.~\ref{stray_comp}). As
seen in Eq.~(\ref{eq_psf}), the PSF and the (local) stray light are mutually
related to each other in a way that makes a separation between them
difficult. To cross-check that the approach of using a single profile averaged
over the full FOV instead of the explicit local version given by the PSF is
reasonable, we calculated the local stray light using the gain term of
Eq.~(\ref{eq_1stcorr}) for all pixels of three data sets: the limb (@396\,nm)
and sunspot (@630\,nm) observations used in the previous section, and a
large-area map taken in quiet Sun (QS) on disk center (@396\,nm, not shown
  here). We then determined the intensity of the local stray-light profiles
relative to the average profile of the full FOV because the number can be used as
first-order proxy\footnote{In Eq.~(\ref{eq_1stcorr}), $I_{\rm obs}$ is used because $I_{\rm true}$ is unknown; $\alpha$ is therefore slightly
  overestimated.}of the true $\alpha$. Figure \ref{hist_alpha} shows the
histograms of the relative frequency of occurrence of a given value of
$\alpha$. In QS regions, $\alpha$ is about 24\,\%. For the limb and the
sunspot data sets it is also generally at or above 20\,\%; however,
it can in some instances fall below 15\,\% (see the tail of the corresponding distributions in the figure). Throughout the umbra of the large sunspot of Fig.~\ref{spot_2d}, $\alpha$ was about 8-10\,\%. 
\section{Discussion \label{sect_disc}}
The various stray-light measurements are basically consistent with the formulation of the stray-light problem in Eqs.~(\ref{eq_stray}), (\ref{eq_psf}), and (\ref{win_loss}). For the special case of POLIS, the stray light contains a significant contribution of spectrally undispersed light created inside the instrument itself that could be absent in other spectrographs. For the spectrally resolved stray light that is created by scattering in the light path upfront of the
grating, we found a lower limit of about 10\,\% from umbral profiles and a
value of about 20\,\% in the quiet Sun, consistently from a limb data set and
an application of the PSF to calculate the local stray light. A
  deconvolution with the instrumental PSF alone, however, does not provide a
good correction for off-limb spectra, presumably because it only covers
the optics behind the telescope focus and includes neither the
  atmospheric scattering nor the telescope PSF of the optics upstream of the focal plane. For off-limb data, additional corrections are needed that have eventually to be adjusted to the specific data set used, but a residual stray-light level below 0.5\,\% can be achieved. For our data, we used the observed intensity variation across a sharp edge with only a minor ad-hoc adjustment to satisfactorily remove the off-limb stray light.

We determined the PSF from observations with a partly blocked FOV, for the two
channels of POLIS and one observational run with the main spectrograph of the
VTT at 777\,nm. For the latter and the 630\,nm channel of POLIS, the tail
region of the stray light far away from the blocking edge was only partially reproduced, i.e., the derived kernel yields less stray light than actually observed. This is possibly related to our basically 1D approach using individual slit spectra. A spatial scan of the blocking edge and the
determination of a 2D PSF may improve the result in the tail
region because in the 1D case only the contamination along the slit is
included, and not the ``lateral'' stray light. Determination of the
  instrumental PSF from observations with a half-blocked FOV is not limited to
  slit-spectrograph data but can equally be done for all kind of 1D or
  2D instruments. It partly allows one to avoid having to use purely ad-hoc constructed PSFs as in \citet{pereira+etal2009}, \citet{joshi+etal2011}, or \citet[][]{rutten+etal2011}. The derivation of the PSF from explicit observations is more solid than the indirect method of a comparison between observed and synthetic spectra from simulations \citep[e.g.,][]{scharmer+etal2011}, where spatial {\em and} spectral resolution effects (cf.~Sect.~\ref{wo_meas}) can get mixed up.

The instrumental PSF can be used to deconvolve the observed spectra post-facto
for both a correction of the stray light and an improvement of the spatial
resolution. We tested the deconvolution on three different data sets, two
spectroscopic and one spectropolarimetric observation. From the visual
impression, the spatial resolution has improved in all cases beyond a simple
increase in contrast owing to the global reduction of intensity because
of the subtraction of the stray-light contribution. The noise level was
increased by the deconvolution by up to a factor of about 2, while our PSF
seems to be well behaved in general with respect to the noise amplification,
presumably because it is sampling-limited in all cases and does not
cover the high spatial frequencies down to the theoretically
  achievable diffraction limit where the noise contribution to the Fourier
  power also becomes important. For the polarimetric observation, we point out that we used the instrumental PSF to deconvolve an observation taken five years before the PSF measurement. Even if this still led to a clear improvement, the PSF measurement should be done closer in time. Before such a deconvolved data set can, however, be used for a scientific investigation, the effects of the deconvolution on the physically relevant quantities like the net circular polarization or the LOS velocities would have to be investigated in more detail \citep[see, e.g.,][]{puschmann+beck2011}. This should include tests with synthetic observations that are convolved with the known PSF. 

There are, however, more reasons that the deconvolution works well
with the data sets used. The PSF is sharply peaked, therefore pixels with a
separation of more than about 2\,--\,3$^{\prime\prime}$ from a spatial
location do not enter strongly. This automatically restricts the correction to
the close surroundings of a given pixel, which is necessary for
slit-spectrograph data with their sequential spatial scanning over a
permanently evolving solar scene. A second reason is that all observations
were obtained using AO correction, therefore the time-variant part of the
instantaneous PSF is already largely compensated for. The third
reason is that all observations initially have a high signal-to-noise
(S/N) ratio because of the comparably long integration times of ten
seconds or more. Finally, the observations were also spatially oversampled for
the data used here; i.e., the spatial sampling along and perpendicular to the slit was below the effective resolution by a factor of two or more.
\section{Conclusions\label{concl}}
Observations with a partly blocked FOV in the first accessible
focal plane allow one to estimate the (static) instrumental
PSF downstream of the focal plane, in case no observations of a planetary
transit are available,  but it seems generally to be recommended to use a 2D
FOV to determine the PSF. The PSF obtained allows one to determine the
local stray-light contamination of a pixel from its close surroundings. For a
homogeneously lit area -- as is common in the quiet Sun -- the explicit calculation by the PSF can be exchanged by an average stray-light contribution proportional to the average profile of the full FOV; in the case of sunspot observations, the stray-light correction should be calculated explicitly using the PSF because of the large intensity gradients. The PSF and the corresponding stray-light level can be cross-checked with observations near the solar limb, where continuum wavelength windows provide an excellent intensity reference. For ground-based observations taken with AO correction at the VTT, with a field stop before the first focal plane, the instrumental PSF seems to be dominating over the contributions from the telescope and the Earth's atmosphere.

The PSF can also be used for a spatial deconvolution using a Fourier
method. This seems a valid option for correcting slit-spectrograph observations post-facto for the static contributions of the instrumental PSF.
\begin{acknowledgements}
The VTT is operated by the Kiepenheuer-Institut f\"ur Sonnenphysik (KIS) at
the Spanish Observatorio del Teide of the Instituto de Astrof\'{\i}sica de
Canarias (IAC). The POLIS instrument has been a joint development of the High
Altitude Observatory (Boulder, USA) and the KIS. C.B.~acknowledges partial
support by the  Spanish Ministerio de Ciencia e Innovaci\'on through projects AYA 2007-63881, AYA 2010-18029, and JCI 2009-04504. D.F.~gratefully acknowledges financial support by the European Commission through the SOLAIRE Network (MTRN-CT-2006-035484) and by the Programa de Acceso a Grandes Instalaciones Cient\'ificas financed by the Spanish Ministerio de Ciencia e Innovaci\'on. We thank C.~Allende Prieto, B.~Ruiz Cobo, M.~Collados, and J.~A.~Bonet for helpful
discussions. We thank the anonymous referee for providing helpful comments.
\end{acknowledgements}
\bibliographystyle{aa}
\bibliography{references_luis_mod}
\begin{appendix}
\section{Derivation of first-order correction\label{appa}}
Solving Eq.~(\ref{win_loss1}) for $I_{\rm  true}$ yields at first the recursive equation
\begin{eqnarray}
I_{\rm  true}(x,y)=\frac{1}{(1-\alpha)}\left( I_{\rm obs}(x,y) -
  \sum_{x^\prime,y^\prime} K({\bf r}) I_{\rm true}(x^\prime,y^\prime) \right) \;,\label{eqq}
\end{eqnarray}
where ${\bf r} = (x-x^\prime,y-y^\prime)$. 

Exchanging the term $I_{\rm  true}(x^\prime,y^\prime)$ with the relation given by
Eq.~(\ref{eqq}) gives
\begin{eqnarray}
I_{\rm true}(x,y) &&=\frac{1}{(1-\alpha)} \left( I_{\rm obs}(x,y) - \sum_{x^\prime,y^\prime} K({\bf r}) \frac{1}{(1-\alpha)} I_{\rm  obs}(x^\prime,y^\prime)- \right.\nonumber\\
&&-\left.\sum_{x^\prime,y^\prime} K({\bf r})
  \sum_{x^{\prime\prime},y^{\prime\prime}} K({\bf
    r}^\prime)\frac{1}{(1-\alpha)} I_{\rm  true}(x^{\prime\prime},y^{\prime\prime})
\right) \,.
\end{eqnarray}
With $\frac{1}{(1-\alpha)} \sim 1 + \alpha$ for $\alpha \ll 1$  and neglecting
all terms $\propto \alpha,\alpha^2, K\,\alpha, K^2$, one obtains
\begin{eqnarray}
I_{\rm  true}(x,y) =  I_{\rm  obs}(x,y) - \sum_{x^\prime,y^\prime}
  K(x-x^\prime,y-y^\prime) I_{\rm  obs}(x^\prime,y^\prime)  \,.
\end{eqnarray}
\section{Convolution kernels at 777\,nm and 630\,nm\label{appb}}
\begin{figure}
\centerline{\resizebox{8.8cm}{!}{\includegraphics{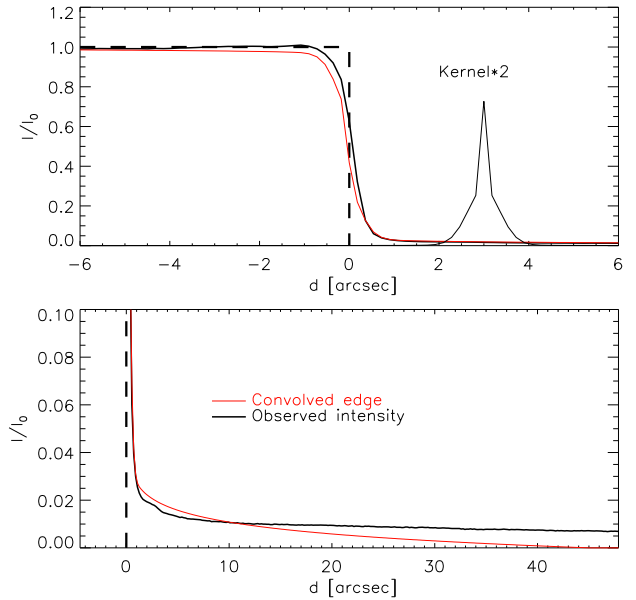}}}
\caption{Derivation of the PSF at 777\,nm. {\em Thick black}: observed
  intensity. {\em Dashed}: edge function. {\em Thin red}: convolved edge.
  {\em Upper panel}: observed intensity across the blocked edge. The
  convolution kernel is shown on the right; for better visibility it was multiplied by 2. {\em Lower panel}: ``tail'' region. \label{777_kernel}}
\end{figure}

\paragraph{Convolution kernel at 777\,nm}
Figure \ref{777_kernel} shows the observed intensity across the location
  of the blocking edge at 777\,nm, together with the convolution kernel used
  to match the convolved edge function to the observation ({\em upper
    panel}). The kernel has more extended wings than the one of POLIS, but we
  were still not able to perfectly reproduce the tail region for distances
  over about 15$^{\prime\prime}$ from the location of the edge ({\em lower panel}). Further increasing the wing contribution led to a mismatch on the other side of the edge in the lit area because the intensity of the convolved edge curve gets reduced more. Up to about 10$^{\prime\prime}$ distance from the edge, the observations are reproduced well by the convolved edge.
\begin{figure}
\centerline{\resizebox{8.8cm}{!}{\includegraphics{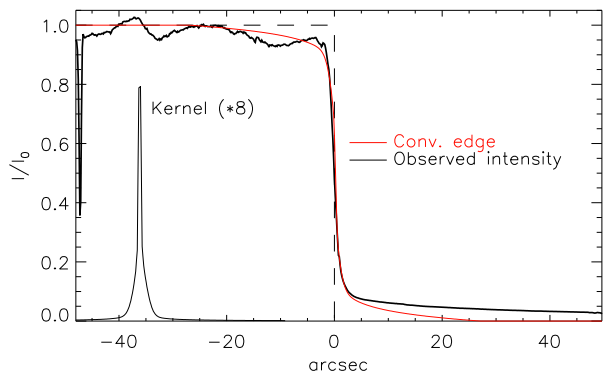}}}
\caption{Test of the PSF at 630\,nm. {\em Thick black}: observed
  intensity. {\em Dashed}: edge function. {\em Thin red}:  convolved edge. The
  convolution kernel is shown on the left; it was multiplied by 8 for better visibility. \label{630_kernel}}
\end{figure}
\paragraph{Convolution kernel at 630\,nm}
For the 630\,nm channel of POLIS, we resampled the kernel
  determined at 396\,nm to the two times finer spatial sampling in the red
  channel by interpolation. Figure \ref{630_kernel} shows that an application of this kernel to
  the edge function reproduces the intensity in the 630\,nm channel for the
  observation with a half-blocked FOV satisfactorily, even if in the tail
  region the match is worse than for 396\,nm. Because of the resampling, the
  amplitude of the kernel is halved compared to Fig.~\ref{fig5}.
\end{appendix}
\end{document}